\documentclass[iop,revtex4]{emulateapj}

\usepackage{natbib}
\usepackage{longtable}
\usepackage{rotating}
\usepackage{epstopdf}
\usepackage{graphicx}
\usepackage{subfigure}
\usepackage[usenames,dvipsnames]{color}
\usepackage{ifthen}
\usepackage[usenames,dvipsnames]{pstricks}
\usepackage{epsfig}
\usepackage{ulem}

\def\cm2{cm$^{-2}$}

\def\nh3{NH$_3$}
\def\n2h{N$_2$H$^+$}

\def\13co{$^{13}$CO}
\def\c18o{C$^{18}$O}
\def\hc3n{HC$_3$N}
\def\h2{H$_2$}
\def\nh{n(H$_2$)}

\def\rsun{\ifmmode {\rm R}_{\mathord\odot}\else $R_{\mathord\odot}$\fi}
\def\msun{\ifmmode {\rm M}_{\mathord\odot}\else $M_{\mathord\odot}$\fi}
\def\lsun{\ifmmode {\rm L}_{\mathord\odot}\else $L_{\mathord\odot}$\fi}

\def\QL#1{{\bf{#1}}}

\begin{document}

\slugcomment{}

\shorttitle{Turbulence Dissipation in Taurus} \shortauthors{Qian, et al.}

\title{Studies of Turbulence Dissipation in Taurus Molecular Cloud with Core Velocity Dispersion (CVD)}

\author{ Lei Qian \altaffilmark{1} \altaffilmark{2}\footnote{LQ carried out the analysis and wrote the paper.}, Di Li \altaffilmark{1} \altaffilmark{2} \altaffilmark{3} \footnote{DL proposed the original CVD method and contributed to the text.}, Yang Gao\altaffilmark{4} \altaffilmark{5}\footnote{YG help put the analysis in astrophysics context and contributed to the text. Now at School of Physics and Astronomy, Sun Yat-Sen University, Zhuhai, China}, Haitao Xu\altaffilmark{4} \altaffilmark{6}\footnote{HTX helped put refine the turbulence analysis and contributed to the text.}, Zhichen Pan \altaffilmark{1} \altaffilmark{2}\footnote{ZCP contributed to the text.}}
\affil{} \altaffiltext{1} {National Astronomical Observatories,
Chinese Academy of Sciences, Beijing, 100012, China}
\altaffiltext{2} {CAS Key Laboratory of FAST, NAOC, Chinese Academy of Sciences}
\altaffiltext{3} {University of Chinese Academy of Sciences, Beijing, China}
\altaffiltext{4} {Center for Combustion Energy, Tsinghua University, Beijing, 100084, China}
\altaffiltext{5} {Department of Physics and Astronomy, Sun Yat-Sen University, Zhuhai, Guangdong 519082, China }
\altaffiltext{6} {School of Aerospace Engineering, Tsinghua University, Beijing, 100084, China}

\begin{abstract}
Turbulence dissipation is an important process affecting the energy balance in molecular clouds, the birth place of stars. Previously, the rate of turbulence dissipation is often estimated with semi-analytic formulae from simulation. Recently we developed a data analysis technique called core-velocity-dispersion (CVD), which, for the first time, provides direct measurements of the turbulence dissipation rate in Taurus, a star forming cloud. {The thus measured dissipation rate of $(0.45\pm 0.05)\times 10^{33}~{\rm erg~s^{-1}}$ is
similar to those from dimensional analysis and also consistent
with the previous energy injection rate based on molecular outflows and bubbles.}
\end{abstract}

\keywords{ISM: clouds --- ISM: molecules --- ISM: individual(Taurus) }

\section{Introduction}
\label{sec:intro}

In molecular clouds, turbulence is a ubiquitous process playing a crucial role in the star formation \citep{2004ARA&A..42..211E,2007ARA&A..45..565M}. {Although turbulence can generate high-density structures and thus enhance the effect of gravity in local and relatively small scales, it is generally  treated as a pressure term, which counteracts gravity, retarding cloud cores from collapsing to form stars. In regions with strong apparent turbulence, such as those near the Galactic center, the star formation efficiency is clearly damped~\citep{2017A&A...603A..89K}.}
Gas cores with comparable gravitational energy and turbulence energy can also form stars after the latter is dissipated if there
  is no continuous turbulence energy injection \citep{Gao2015}.
Therefore, turbulence energy dissipation rate is a key parameter to determine the time scale of star formation.

Turbulence energy can be injected by differential rotation of galactic disk \citep{1981ApJ...246L.151F}, galactic disk tidal
  force \citep{2015MNRAS.446..973F}, large-scale gravitational instabilities in galactic disks \citep{2003ApJ...590..271E,2010MNRAS.409.1088B}, stellar feedback \citep{2012ApJ...752..146L}, supernova explosions \citep{2005A&A...436..585D,2009ApJ...704..137J,2016ApJ...822...11P}, and fluctuations in Galactic synchrotron radiation \citep{2017MNRAS.466.2272H}.
The injected energy will cascade down to small scales and dissipate through viscous processes (in this case at Kolmogorov scale)
  or low velocity shocks \citep{2012ApJ...748...25P}.
The dissipation of turbulence energy evolves from viscosity dominated ($\mathcal{M}_{\rm s}\lesssim 1$)
  to shock dominated ($\mathcal{M}_{\rm s}\gtrsim 10$) regimes with increasing rms sonic Mach number $\mathcal{M}_{\rm s}$ \citep{2004PhRvL..92s1102P}
  \footnote{{In the weakly ionized interstellar media like molecular clouds, the dominant dissipation mechanism of MHD turbulence is ion-neutral collisional damping, i.e. ambipolar diffusion \citep{Lithwick2001,2016ApJ...833..215X}.}}.
In a typical molecular cloud, such as Taurus, $1\lesssim \mathcal{M}_{\rm s}\lesssim 10$, so both dissipation mechanisms could take effect.
The typical Kolmogorov scale in molecular clouds is estimated to be  $10^{-5}\sim 10^{-4}$ pc, considering thermal viscosity
  \citep{2007ARA&A..45..565M,2011ApJ...737...13K}.
There have been some unsuccessful attempts to probe these scales with the slope change of the turbulence energy spectrum \citep{2008ApJ...677.1151L}.

It is conceivable, but rarely attempted, to measure the turbulence dissipation rate by observing the excess emission from gas,
  presumably excited by turbulence.
Goldsmith et al.\ (2010) detected NIR H$_2$ emission across the boundary of the Taurus molecular cloud.
Since H$_2$ transitions are hundreds of Kelvins above the ground state,
  such emission remains a mystery, given the much lower temperature and the lack of UV source in Taurus.
One possibility is excitation by shocks.
Recent studies suggest that the shock induced turbulence energy dissipation can be traced by mid-$J$ (e.g. $J$=6-5) CO lines \citep{2014MNRAS.445.1508P}.
However, neither NIR H$_2$ emission nor mid-$J$ CO lines can be readily observed to trace viscous dissipation.

The turbulence dissipation rate $\dot{E}_{\rm diss}$ in molecular clouds was also estimated with semi-analytical formulae based on numerical simulations
  \cite{Mac_Low1999}.
Such estimate is in essence equivalent to dimensional analysis \citep{2007ARA&A..45..565M}.
The key parameter in these analysis is the turbulence dissipation time, which is on the order of the turbulence crossing time.
For Taurus clouds, these methods gave an estimate $\dot{E}_{\rm diss} \sim$ $0.7\times 10^{33}-3.8\times 10^{33}~{\rm erg~s^{-1}}$
  \citep{Li2015,2012MNRAS.425.2641N}.

There are other observation-based ways to estimate the turbulence energy dissipation rate, e.g., with structure functions \citep{1995tlan.book.....F}.
The calculation of structure functions needs three-dimensional position and velocity, rarely available in astronomical observations.
We have developed a new method, namely the core-velocity-dispersion (CVD) method, to study the cloud and turbulence structures \citep{Qian2012}.
In the present work, we developed a CVD-based method to estimate the turbulence dissipation rate $\dot{E}_{\rm diss}$ in a thin and face-on cloud.
As an example, the turbulence energy dissipation rate in Taurus molecular cloud was estimated.

\section{Methods}
\label{sec:methods}

\subsection{Structure Function and CVD}
{In astrophysical observations, since the celestial objects are projected onto a 2D surface, the structure functions are hard to measure directly.
In a laboratory setting, point-like objects can be placed into
the turbulent flow as tracers of the flow motion \citep{2001Natur.409.1017L}.
In molecular clouds, although there is no way to place artificial objects. There exist, however, condensed and well-localized
objects in the molecular cloud, namely, the molecular cores. According to cores' generally accepted definition, a core has volume density more than a order-of-magnitude higher than its surroundings and occupy only a small fraction of the total cloud volume~\citep{2007prpl.conf...33W}.  Although not as discrete from the large-scale flow as artificial objects, cores still reveal characteristics of its ambient turbulence.}

{We developed a dynamic analysis tool, namely Core-Velocity-Dispersion (CVD) based on the collective motion of each core.
Utilizing the peak line-of-sight velocity (redshift) of the emission profiles of molecular cores, \cite{Qian2012} found that the velocity difference between each pair of cores for a certain spatial scale (i.e. CVD) to depend on the projected distance between cores. In Taurus, the relation between CVD and core distance follows the same trend, namely Larson's law \citep{Larson1981}, as that between the width of the molecular line and the size of the clouds. It necessarily follows that the cores traces the general turbulence flow just like point-like tracers in laboratories and that Taurus is a nearly face-on thin cloud \citep{Qian2015}.}

In general, turbulence energy cascades from large (injection) to small (dissipation) scales. The rate of energy cascade equals the rate of energy dissipation in a statistically stationary turbulent flow. At relatively large scales in Taurus molecular cloud, the turbulence in molecular cloud can be approximated as incompressible as shown in section \ref{sec:results}. For incompressible turbulence in
the inertial range, the rate of energy cascade per unit mass, $\epsilon$, is related to the second order longitudinal and transverse structure functions, $S^2_{\rm ll}$ and $S^2_{\rm tt}$, as \cite{Antonia1997}
\begin{equation}
S^2_{\rm ll}=C_{12}\epsilon^{2/3} l^{2/3}
\label{sll}
\end{equation}
and
\begin{equation}
S^2_{\rm tt}=\frac{8}{3}C_{12}\epsilon^{2/3} l^{2/3},
\label{stt}
\end{equation}
where $C_{12}=2.12$ is a universal constant \cite{Turbulent} and $l$ is the length scale.
As defined by \cite{Kolmogorov1991}, the structure functions
\begin{equation}
S^2_{\rm ll}(l_{12})=\left\langle \left(v_{\rm l2}-v_{\rm l1}\right)^2\right\rangle=\left\langle \delta v_{\rm l}^2\right\rangle
\end{equation}
and
\begin{equation}
S^2_{\rm tt}(l_{12})=\left\langle \left(v_{\rm t2}-v_{\rm t1}\right)^2\right\rangle=\left\langle \delta v_{\rm t}^2\right\rangle,
\label{Stt2}
\end{equation}
where the length $l_{12}$  is a line segment $C_1C_2$  in a turbulence flow, and $v_{\rm l}$ and $v_{\rm t}$ are the velocity components along and perpendicular to $C_1C_2$.
  By measuring the energy cascading rate, the energy dissipation rate can be obtained, as long as the turbulent energy finally dissipates at some small scales. The transverse and the longitudinal structure function are related as  \cite{Turbulent}
\begin{equation}
S^2_{\rm tt}=2 \left(1+\frac{l}{2}\frac{\partial}{\partial l}\right)S^2_{\rm ll}.
\end{equation}

\begin{figure}[htbp]
\centering
\includegraphics[width=8cm]{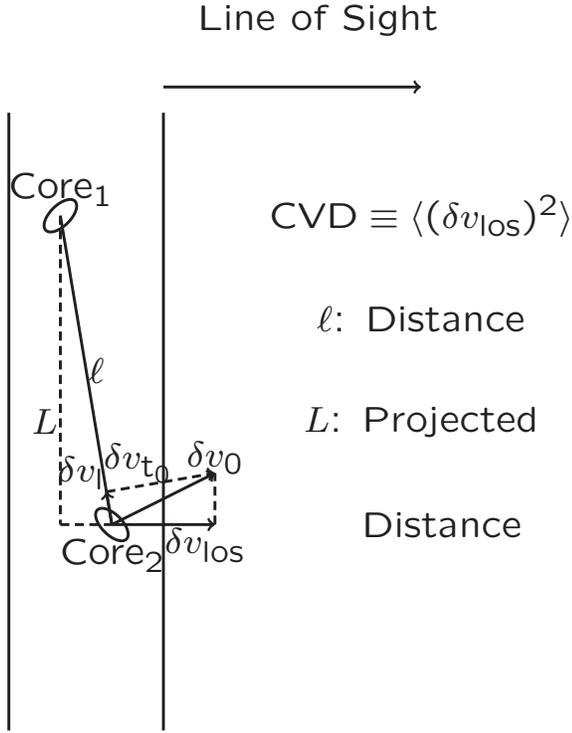}
\caption{Sketch of the core velocity dispersion (CVD). CVD$\equiv \langle (\delta v_{\rm los})^2\rangle^{1/2}$, where $\delta v_0$ is the full velocity difference of the core pair, and $\delta v_{los}$ is the line-of-sight component of $\delta v_0$. $\delta v_{\rm l}$ and $\delta v_{\rm t0}$ are the longitudinal velocity difference, and the difference of the transverse  velocity component contributing to the line of sight velocity. }
\label{cartoon}
\end{figure}

When the cloud is thin, i.e.\ $h\ll L$, where $h$ is the thickness and $L$ the transverse scale, the transverse structure function can be obtained from CVD, since the projected distance is equivalent to the 3D distance ($l\sim \sqrt{L^2+h^2}\sim L$).
CVD is defined as CVD$\equiv \langle \delta v_{\rm los}^2\rangle^{1/2}$ (where $v_{\rm los}$ is the line of sight velocity, see Fig.~\ref{cartoon}). For a cloud with a finite thickness, the line of sight velocity component has contributions from the longitudinal velocity, so the difference of the line of sight velocity $\delta v_{\rm los}=\sin\theta\delta v_{\rm t_0}+\cos\theta\delta v_{\rm l}$, where $\theta$ is the angle between the line of sight and the longitudinal direction  and $\delta v_{\rm t_0}$ is the transverse
component of the velocity difference that contributes to the line of sight velocity difference. Since $\langle \delta v_{\rm t_0} \delta v_{\rm l}\rangle=0$, $\langle\delta {v_{\rm t_0}}^2\rangle=\frac{1}{2}S_{\rm tt}^2$, $\langle\delta v_{\rm l}^2\rangle=S^2_{\rm ll}=\frac{3}{8}S_{\rm tt}^2$, we have CVD$^2=\langle (\delta v_{\rm los})^2\rangle=\langle (\sin\theta\delta v_{\rm t_0}+\cos\theta\delta v_{\rm l})^2\rangle=\frac{1}{2}
S_{\rm tt}^2\left(\langle\sin^2\theta\rangle+\frac{3}{4}\langle\cos^2\theta\rangle\right)$.
Define
$f \equiv \langle\sin^2\theta\rangle +\frac{3}{4}\langle\cos^2\theta\rangle$.
So CVD is related to the transverse structure function as
\begin{equation}
{\rm CVD}^2\equiv \left\langle \delta v_{\rm los}^2\right\rangle = \frac{1}{2}f S^2_{tt}.
\label{CVD}
\end{equation}
Combining equations \ref{stt} and \ref{CVD}, we get
\begin{equation}
\epsilon=\frac{1}{L}\left(\frac{{\rm CVD}^2}{\frac{4}{3}C_{12}f}\right)^{3/2},
\label{dissipation_rate}
\end{equation}
where the energy cascade rate $\epsilon$ is related to an observable, CVD, plus a geometrical factor $f$. In the extreme cases of $L\gg h$ and $L\ll h$, $f\approx$  $1$ and \QL{$\frac{3}{4}$}, respectively.

\subsection{Estimate of $S^2_{\rm tt}/\rm CVD^2$}

\begin{figure}[htbp]
\centering
\includegraphics[width=8cm]{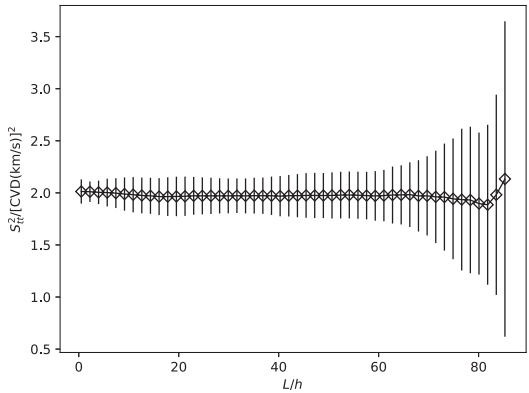}
\caption{Ratio of $S^2_{\rm tt}$ and CVD$^2$ obtained by fractional Brownian motion modeling. The horizontal axis is the ratio of projected scale to the thickness $L/h$. $S^2_{\rm tt}/{\rm CVD}^2= 2/f\approx 2.0\pm 0.2$. The error gets larger at larger scales because the number of core pairs get smaller at large scale.
}\label{sttcvd2_ratio}
\end{figure}

We used the fractional Brownian motion model  to estimate the ratio $S^2_{\rm tt}/\rm CVD^2$ $(=2/f)$, following the procedures described below.  First, we generate a random Gaussian velocity field on a 3D grid. Second, we perform a Fourier transform to get a field in frequency domain ($k$-space, $k$ is the wave number). Third, we process this field in the frequency domain to satisfy the desired power law energy spectrum \citep[e.g.,\ $E(k)\propto k^{-5/3}$, ][]{Qian2015}. Fourth, we perform an inverse Fourier transform and normalize the generated field to fulfill the desired variance and its dependence on the velocity dispersion.
These randomly generated cores are used to calculate both $S^2_{\rm tt}$ and CVD$^2$. As shown in figure~\ref{sttcvd2_ratio}, $S^2_{\rm tt}/{\rm CVD}^2= 2/f\approx 2.0\pm 0.2$ at $L/h\lesssim 50$. So $f\approx 1.0 $, consistent with the estimates in the previous subsection. The error gets larger at larger scales due to a lack of sampling of core pairs when the distance between cores gets close to the size of the map. We use $f=1.0$ and $\Delta f=0.1$ for the following calculations of the turbulence energy dissipation rate and its uncertainty.

\begin{figure}[htbp]
\centering
\includegraphics[width=8cm]{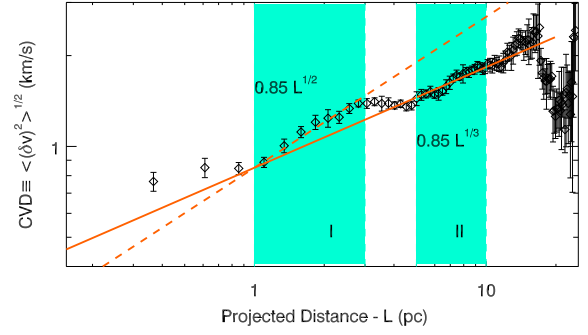}
\caption{CVD - Projected Distance relation for Taurus cloud. The dashed line and the solid line are $0.85 L^{1/2}$ and $0.85 L^{1/3}$, respectively. The data between 1 pc and 3 pc (region I) can be fitted with $(0.85\pm 0.01) L^{1/2}$. The data between 5 pc and 10 pc (region II) can be fitted with $(0.85\pm 0.003) L^{1/3}$.}\label{cvd}
\end{figure}

\section{Turbulence Dissipation Rate in Taurus}
\label{sec:results}

Taurus is a nearly face-on thin cloud \citep{Li2015,Qian2015}, we quantified the line-of-sight dimension of Taurus to be less than 1/8 of its on-the-sky size.

Such a favorable geometry allows us to use the method described in Section \ref{sec:methods} to probe the dissipation rate.
Between $5\lesssim L\lesssim 10\ \rm pc$, the Taurus CVD was found to follow the $L^{1/3}$ scaling (see Fig.~\ref{cvd}).
Eq.\ \ref{dissipation_rate} then gives the turbulence energy cascade rate
\begin{equation}
\epsilon = (0.15\pm 0.02) \times 10^{-4} \  {\rm erg\ s^{-1}g^{-1}} .
\end{equation}
The uncertainty comes from both $S^2_{\rm tt}/{\rm CVD}^2$ (Fig.~\ref{sttcvd2_ratio}) and CVD (Fig.~\ref{cvd}), but mainly from the former.

The original work by Larson (1981) obtained the Larson's relation $\Delta v\propto L^{\beta}$, with $\beta=0.38$. The subsequent seminal work by Solomon et al.\ (1985) revised the index to be $\beta=0.5$ and attributed the steeper power to the compressible nature of the gas cloud.

In compressible fluid, the velocity fluctuation scales with both the density and the scale. Equivalently, the structure function $S_{\rm tt}\sim (\ell/\rho)^{2/3}$ instead of $\ell^{2/3}$, with $\rho$ being the density, in Eq.~\ref{stt}. Two empirical evidence support our treating the gas as incompressible in the intra-range (5-10 pc) in this work. First, we clearly recovered the original Larson's relation in the intra-range, which is consistent with gas being incompressible. Second, the cores are condensations with much higher density than the ambient gas. By treating cores as point masses with a single line of sight velocity $v_{\rm los}$, CVD is only sensitive to scales beyond the core diameters, which fall in the intra-range. The density of the gas in the intra-range does not vary much. The density of the gas increase significantly only when they condense into cores.

The total mass of Taurus molecular cloud is $M=1.50\times10^4 \ M_{\odot}$~\citep{Pineda2010}. The total turbulence energy dissipation rate of Taurus molecular cloud is then
\begin{equation}
\dot{E}_{\rm diss}=\epsilon M=(0.45\pm 0.05)\times 10^{33}\ {\rm erg/s}.
\end{equation}

In previous studies, the turbulence energy dissipation rate is estimated by dividing the energy from outflows and bubbles with a typical timescale.
 The turbulence energy dissipation rate of Taurus molecular cloud thus estimated is $0.7-3.1\times 10^{33}$ erg/s~\citep{Li2015} and $3.8\times 10^{33}$ erg/s~\citep{2012MNRAS.425.2641N}, which corresponds to $\epsilon=0.2-1.0\times 10^{-4}\ {\rm erg\ s^{-1}g^{-1}}$  and $\epsilon=1.3\times 10^{-4}\ {\rm erg\ s^{-1}g^{-1}}$, respectively (see appendix). Our result based on the CVD method is independent of these dimensional analyses and turns out to be consistent with these estimates, in the sense that the dissipation rate estimated here is smaller/comparable to that estimated from outflows and bubbles.
This consistency provides additional support to our method.

\section{Discussion}
\label{sec:discussion}

We used molecular cores as an approximate and practical tracer of turbulent flow in molecular clouds.
{Although the internal motion of the cores (smaller scales) could be dominated by compressive motions, we do not expect CVD to be
of much bias in this regard, as only the collective properties (peak velocity of the whole core) are being used. }

{There exit other statistical methods for obtaining structure functions (of Faraday rotation measure) in terms of
the projected separation for both thin and thick clouds~\cite[e.g.][]{Lazarian2016}.
These structure functions are calculated with the integrated value (along the line of sight) at each point on the projected plane. On the contrary, CVD is calculated based on collective characteristics of each core, thus having a much better localization property than existing methods. For example, when cores overlap with each other in projected 2D space, CVD can resolve them in the spectral dimension, which was demonstrated in~\cite{Qian2012}.}

\QL{CVD, in the current incarnation, is limited by the lack of knowledge of separation along the line of sight.
Previously, \cite{Qian2015} looked into the effect of using projected distance. The main obvious conclusion is the fact
that as long as there is any correlation existing between CVD and the projected distance, the cloud cannot be thick. In a thick cloud,
the motion of cores at different locations should have no dependence whatsoever on the projected distance between them.
Indeed, we explored this projection effect and quantified the thickness of Taurus to be smaller than the 1/8 of the cloud transverse scale, i.e.,
Taurus is thin! The recipe presented in this paper utilizing CVD thus only works for a thin cloud. }
\QL{As shown in Figure 2, if the thickness of the cloud is still in the inertial range, then the procedure described in the manuscript can still yield accurate results even if the projected distance is comparable to the thickness. However, when the thickness is large, i.e., larger than the large-eddy size, the proportionality between CVD$^2$ and $S_{\rm tt}$ given by Eq.~\ref{CVD} will be destroyed and the method cannot be used, unless detailed information along the line of sight is known.}


The scaling laws in Eq.~\ref{sll} and \ref{stt} are rigorously correct only
for incompressible turbulence.
For compressible turbulence, no simple analytical relation exists.
The fact that the relative motion between dense cores seem to follow the general cloud turbulent flow, i.e.\ CVD
mimics the original Larson's law \cite{Larson1981} allows us to trust the CVD measurement to the degree that it does not deviate from incompressible turbulence by orders of magnitude.

At small scales, the estimates of the structure function $S^2_{\rm tt}$ with CVD will be affected by the finite thickness $h$ of the cloud. The structure function at relatively larger scales $L\sim 5-10 \ \rm pc$ was used to estimate
the turbulence energy dissipation rate (Fig. \ref{cvd}). In this
sense the estimates of the turbulence energy dissipation rate in
this paper does not depend on either the energy injection or the dissipation mechanism, but only relies on the scaling laws of turbulence energy cascade.
This energy dissipation rate derived from the CVD method can be further used to estimate the turbulence decay rate/time in cloud cores \cite{Gao2015}.
In this case how the decay of turbulence in cloud cores facilitates the star formation activity can be qualitatively studied.


It is also interesting to compare the energy dissipation rate with the cloud cooling rate. The cooling rate per H$_2$ molecule is about $10^{-27}\ \rm erg/s$ for a volume density of $10^3\ \rm cm^{-3}$~\cite{1995ApJS..100..132N}. The total cooling rate for Taurus molecular cloud is then $\sim 9.0\times 10^{33}\ \rm erg/s$ for an average volume density of $10^3\ \rm cm^{-3}$. This cooling rate is higher than the turbulence dissipation.

\section{Summary}
\label{sec:summary}

The transverse structure function $S^2_{\rm tt}$ can be estimated with core velocity dispersion (CVD) in a thin and face-on molecular cloud. The ratio $S^2_{\rm tt}/{\rm CVD}^2$ is found to be $2.0\pm 0.2$, based on fractional Brownian motion model.
The measured turbulence energy dissipation rate of $(0.45\pm 0.05)\times 10^{33}~{\rm erg~s^{-1}}$ for scales between 5 and 10 pc matches previous observational estimates. Such a dissipation rate is also consistent with the energy injection rate from star formation feedback at relatively smaller scales between 0.05-0.5 pc \citep{Li2015}. An empirical picture of the turbulence in Taurus molecular cloud is that, the majority of energy injection happens at cloud complex scales ($>10$ pc), the energy then cascades through the intermediate scales down to clump scales while counterbalance the gravity in rough viral equilibrium. It finally reaches a dynamic balance with star formation feedback at small scales of clumps and cores.

\acknowledgments This work is supported by National Key R\&D Program of China
No. 2017YFA0402600, State Key Development Program for Basic Research (2015CB857100), the National Natural Science Foundation of China No. 11373038, No. 11373045, No. 11672157, No. 11735313, the CAS International Partnership Program No.114A11KYSB20160008. LQ is supported in part by the Youth Innovation Promotion Association of CAS (id.~2018075)


\begin{thebibliography}{37}
\expandafter\ifx\csname natexlab\endcsname\relax\def\natexlab#1{#1}\fi

\bibitem[{{Antonia} {et~al.}(1997){Antonia}, {Ould-Rouis}, {Zhu}, \&
  {Anselmet}}]{Antonia1997}
{Antonia}, R.~A., {Ould-Rouis}, M., {Zhu}, Y., \& {Anselmet}, F. 1997, EPL
  (Europhysics Letters), 37, 85

\bibitem[{{Berry} {et~al.}(2013){Berry}, {Reinhold}, {Jenness}, \&
  {Economou}}]{2013ascl.soft11007B}
{Berry}, D.~S., {Reinhold}, K., {Jenness}, T., \& {Economou}, F. 2013, {CUPID:
  Clump Identification and Analysis Package}, Astrophysics Source Code Library

\bibitem[{{Bournaud} {et~al.}(2010){Bournaud}, {Elmegreen}, {Teyssier},
  {Block}, \& {Puerari}}]{2010MNRAS.409.1088B}
{Bournaud}, F., {Elmegreen}, B.~G., {Teyssier}, R., {Block}, D.~L., \&
  {Puerari}, I. 2010, \mnras, 409, 1088

\bibitem[{{de Avillez} \& {Breitschwerdt}(2005)}]{2005A&A...436..585D}
{de Avillez}, M.~A., \& {Breitschwerdt}, D. 2005, \aap, 436, 585

\bibitem[{{Elmegreen} {et~al.}(2003){Elmegreen}, {Elmegreen}, \&
  {Leitner}}]{2003ApJ...590..271E}
{Elmegreen}, B.~G., {Elmegreen}, D.~M., \& {Leitner}, S.~N. 2003, \apj, 590,
  271

\bibitem[{{Elmegreen} \& {Scalo}(2004)}]{2004ARA&A..42..211E}
{Elmegreen}, B.~G., \& {Scalo}, J. 2004, \araa, 42, 211

\bibitem[{{Falceta-Goncalves} {et~al.}(2015){Falceta-Goncalves}, {Bonnell},
  {Kowal}, {L{\'e}pine}, \& {Braga}}]{2015MNRAS.446..973F}
{Falceta-Goncalves}, D., {Bonnell}, I., {Kowal}, G., {L{\'e}pine}, J.~R.~D., \&
  {Braga}, C.~A.~S. 2015, \mnras, 446, 973

\bibitem[{{Fleck}(1981)}]{1981ApJ...246L.151F}
{Fleck}, Jr., R.~C. 1981, ApJL, 246, L151

\bibitem[{{Frisch}(1995)}]{1995tlan.book.....F}
{Frisch}, U. 1995, {Turbulence. The legacy of A. N. Kolmogorov.}

\bibitem[{{Gao} {et~al.}(2015){Gao}, {Xu}, \& {Law}}]{Gao2015}
{Gao}, Y., {Xu}, H., \& {Law}, C.~K. 2015, \apj, 799, 227

\bibitem[{{Herron} {et~al.}(2017){Herron}, {Federrath}, {Gaensler}, {Lewis},
  {McClure-Griffiths}, \& {Burkhart}}]{2017MNRAS.466.2272H}
{Herron}, C.~A., {Federrath}, C., {Gaensler}, B.~M., {Lewis}, G.~F.,
  {McClure-Griffiths}, N.~M., \& {Burkhart}, B. 2017, \mnras, 466, 2272

\bibitem[{{Joung} {et~al.}(2009){Joung}, {Mac Low}, \&
  {Bryan}}]{2009ApJ...704..137J}
{Joung}, M.~R., {Mac Low}, M.-M., \& {Bryan}, G.~L. 2009, \apj, 704, 137

\bibitem[{{Kauffmann} {et~al.}(2017){Kauffmann}, {Pillai}, {Zhang}, {Menten},
  {Goldsmith}, {Lu}, \& {Guzm{\'a}n}}]{2017A&A...603A..89K}
{Kauffmann}, J., {Pillai}, T., {Zhang}, Q., {Menten}, K.~M., {Goldsmith},
  P.~F., {Lu}, X., \& {Guzm{\'a}n}, A.~E. 2017, \aap, 603, A89

\bibitem[{{Kolmogorov}(1991)}]{Kolmogorov1991}
{Kolmogorov}, A.~N. 1991, Proceedings: Mathematical and Physical Sciences, 434,
  15

\bibitem[{{Kritsuk} {et~al.}(2011){Kritsuk}, {Nordlund}, {Collins}, {Padoan},
  {Norman}, {Abel}, {Banerjee}, {Federrath}, {Flock}, {Lee}, {Li},
  {M{\"u}ller}, {Teyssier}, {Ustyugov}, {Vogel}, \& {Xu}}]{2011ApJ...737...13K}
{Kritsuk}, A.~G., {et~al.} 2011, \apj, 737, 13

\bibitem[{{La Porta} {et~al.}(2001){La Porta}, {Voth}, {Crawford}, {Alexander},
  \& {Bodenschatz}}]{2001Natur.409.1017L}
{La Porta}, A., {Voth}, G.~A., {Crawford}, A.~M., {Alexander}, J., \&
  {Bodenschatz}, E. 2001, \nat, 409, 1017

\bibitem[{{Larson}(1981)}]{Larson1981}
{Larson}, R.~B. 1981, \mnras, 194, 809

\bibitem[{{Lazarian} \& {Pogosyan}(2016)}]{Lazarian2016}
{Lazarian}, A., \& {Pogosyan}, D. 2016, \apj, 818, 178

\bibitem[{{Lee} {et~al.}(2012){Lee}, {Murray}, \&
  {Rahman}}]{2012ApJ...752..146L}
{Lee}, E.~J., {Murray}, N., \& {Rahman}, M. 2012, \apj, 752, 146

\bibitem[{{Li} {et~al.}(2015){Li}, {Li}, {Qian}, {Xu}, {Goldsmith},
  {Noriega-Crespo}, {Wu}, {Song}, \& {Nan}}]{Li2015}
{Li}, H., {et~al.} 2015, ApJS, 219, 20

\bibitem[{{Li} \& {Houde}(2008)}]{2008ApJ...677.1151L}
{Li}, H.-b., \& {Houde}, M. 2008, \apj, 677, 1151

\bibitem[{{Lithwick} \& {Goldreich}(2001)}]{Lithwick2001}
{Lithwick}, Y., \& {Goldreich}, P. 2001, \apj, 562, 279

\bibitem[{{Mac Low}(1999)}]{Mac_Low1999}
{Mac Low}, M.-M. 1999, \apj, 524, 169

\bibitem[{{McKee} \& {Ostriker}(2007)}]{2007ARA&A..45..565M}
{McKee}, C.~F., \& {Ostriker}, E.~C. 2007, \araa, 45, 565

\bibitem[{{Narayanan} {et~al.}(2008){Narayanan}, {Heyer}, {Brunt}, {Goldsmith},
  {Snell}, \& {Li}}]{2008ApJS..177..341N}
{Narayanan}, G., {Heyer}, M.~H., {Brunt}, C., {Goldsmith}, P.~F., {Snell}, R.,
  \& {Li}, D. 2008, \apjs, 177, 341

\bibitem[{{Narayanan} {et~al.}(2012){Narayanan}, {Snell}, \&
  {Bemis}}]{2012MNRAS.425.2641N}
{Narayanan}, G., {Snell}, R., \& {Bemis}, A. 2012, \mnras, 425, 2641

\bibitem[{{Neufeld} {et~al.}(1995){Neufeld}, {Lepp}, \&
  {Melnick}}]{1995ApJS..100..132N}
{Neufeld}, D.~A., {Lepp}, S., \& {Melnick}, G.~J. 1995, \apjs, 100, 132

\bibitem[{{Padoan} {et~al.}(2004){Padoan}, {Jimenez}, {Nordlund}, \&
  {Boldyrev}}]{2004PhRvL..92s1102P}
{Padoan}, P., {Jimenez}, R., {Nordlund}, {\AA}., \& {Boldyrev}, S. 2004,
  Physical Review Letters, 92, 191102

\bibitem[{{Padoan} {et~al.}(2016){Padoan}, {Pan}, {Haugb{\o}lle}, \&
  {Nordlund}}]{2016ApJ...822...11P}
{Padoan}, P., {Pan}, L., {Haugb{\o}lle}, T., \& {Nordlund}, {\AA}. 2016, \apj,
  822, 11

\bibitem[{{Pineda} {et~al.}(2010){Pineda}, {Goldsmith}, {Chapman}, {Snell},
  {Li}, {Cambr{\'e}sy}, \& {Brunt}}]{Pineda2010}
{Pineda}, J.~L., {Goldsmith}, P.~F., {Chapman}, N., {Snell}, R.~L., {Li}, D.,
  {Cambr{\'e}sy}, L., \& {Brunt}, C. 2010, ApJ, 721, 686

\bibitem[{{Pon} {et~al.}(2012){Pon}, {Johnstone}, \&
  {Kaufman}}]{2012ApJ...748...25P}
{Pon}, A., {Johnstone}, D., \& {Kaufman}, M.~J. 2012, \apj, 748, 25

\bibitem[{{Pon} {et~al.}(2014){Pon}, {Johnstone}, {Kaufman}, {Caselli}, \&
  {Plume}}]{2014MNRAS.445.1508P}
{Pon}, A., {Johnstone}, D., {Kaufman}, M.~J., {Caselli}, P., \& {Plume}, R.
  2014, \mnras, 445, 1508

\bibitem[{{Pope}(2000)}]{Turbulent}
{Pope}, S.~B. 2000, {Turbulent Flows}, ed. {Pope, S. B.} (Cambridge University
  Press)

\bibitem[{{Qian} {et~al.}(2012){Qian}, {Li}, \& {Goldsmith}}]{Qian2012}
{Qian}, L., {Li}, D., \& {Goldsmith}, P.~F. 2012, ApJ, 760, 147

\bibitem[{{Qian} {et~al.}(2015){Qian}, {Li}, {Offner}, \& {Pan}}]{Qian2015}
{Qian}, L., {Li}, D., {Offner}, S., \& {Pan}, Z. 2015, \apj, 811, 71

\bibitem[{{Ward-Thompson} {et~al.}(2007){Ward-Thompson}, {Andr{\'e}},
  {Crutcher}, {Johnstone}, {Onishi}, \& {Wilson}}]{2007prpl.conf...33W}
{Ward-Thompson}, D., {Andr{\'e}}, P., {Crutcher}, R., {Johnstone}, D.,
  {Onishi}, T., \& {Wilson}, C. 2007, Protostars and Planets V, 33

\bibitem[{{Xu} \& {Lazarian}(2016)}]{2016ApJ...833..215X}
{Xu}, S., \& {Lazarian}, A. 2016, \apj, 833, 215

\end{thebibliography}

\appendix

\section{Data and $^{13}$CO Cores}
\label{sec:data}
For a contiguous spectral survey of any nearby star forming clouds, the FCRAO Taurus survey (Goldsmith et al.\ 2008)  boasts of the best spatial dynamic range (linear size / resolution), which makes it an ideal data set to study turbulence in molecular interstellar medium. The $^{13}$CO (J=1-0, 110.2014 GHz) data of this survey were obtained with the 13.7 m FCRAO telescope between 2003 and 2005. The map is centered at ${\rm RA}(2000.0)=04^{\rm h} 32^{\rm m} 44.6^{\rm s}$, ${\rm Dec}(2000.0)=24^\circ 25' 13.08"$, with an area of $\sim 98\ \rm deg^2$, a spatial resolution of $\sim 45''$, the velocity resolution of 0.266 km/s, and a noise level of 0.1 K  \cite{2008ApJS..177..341N}.

{$^{13}$CO cores are defined as Gaussian components in the $^{13}$CO data cube (p-p-v cube)~\citep{Qian2012}. We used the GAUSSCLUMPS method in the Starlink software {package CUPID~\citep{2013ascl.soft11007B} } to identify cores from the data cube. The lower thresholds of the peak intensity of cores were set to be 7 times the noise level, which is about 0.7 K}. 588 relevant cores were identified and used to calculate the CVD. Each core, whose centroid velocity is used, serves as a sampling point of the turbulence velocity field. The typical size of a core is $\sim 0.1\  \rm pc$. 
For CVD analysis, the core pairs with distance significantly larger than typical core size produce more reliable measurements.

\section{Comparison with other works}

 \begin{figure}[htbp]
 \centering
 \includegraphics[width=16cm]{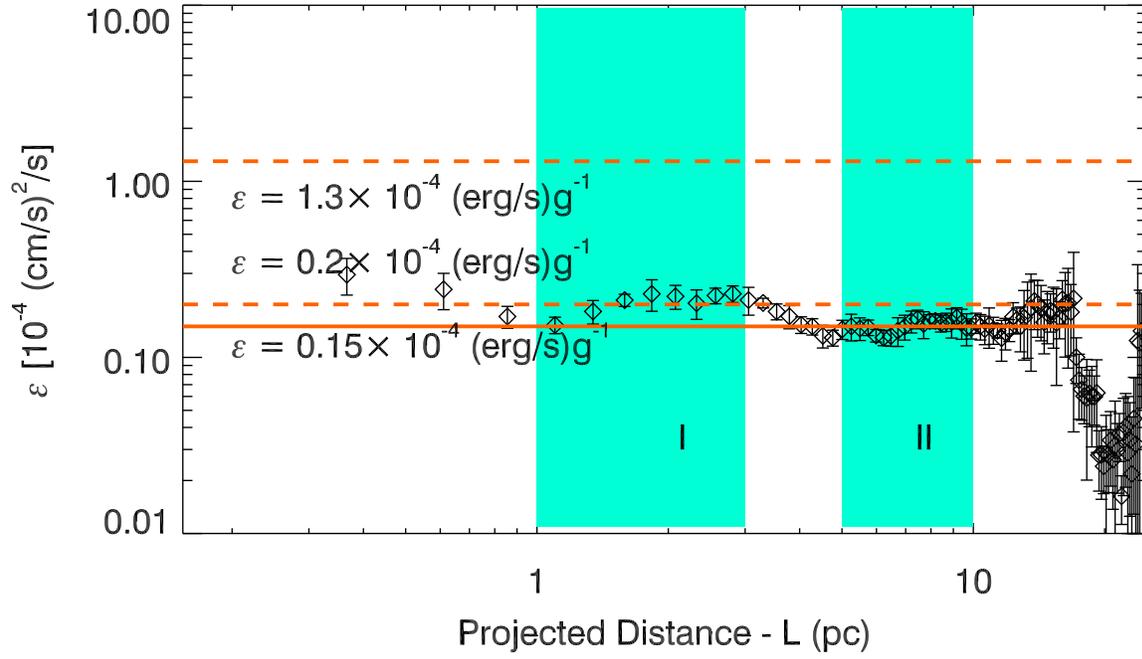}
 \caption{The dissipation rate estimated with CVD. The solid line indicate the dissipation rate estimated with the data between 5 pc and 10 pc. The dashed line shows the dissipation rate obtained from the literature \citep{Li2015,2012MNRAS.425.2641N}. }\label{epsilon}
 \end{figure}

\end{document}